\documentclass[aps,preprint]{revtex4}

\usepackage{epsf}

\begin{document}

\title{{\it Ab Initio} Treatments of the Ising Model in a Transverse 
  Field}

\author{R.F. Bishop,$^{1}$ D.J.J. Farnell,$^{1}$ and
  M.L. Ristig$^{2}$}

\vskip 10 truept

\affiliation{$^1$Department of Physics, University of Manchester Institute of
  Science and Technology (UMIST), P O Box 88, Manchester M60 1QD, United 
  Kingdom.}

\affiliation{$^2$Institut f\"ur Theoretische Physik, 
  Universit\"at zu K\"oln, Z\"ulpicher Str., 50937 K\"oln, Germany.}

\vskip 10 truept

\begin{abstract}
In this article, new results are presented for the 
zero-temperature ground-state properties of the spin-half 
transverse Ising model on various lattices using three 
different approximate techniques. These are, respectively, the coupled cluster 
method, the correlated basis function method, 
and the variational quantum Monte Carlo method. 
The methods, at different levels of approximation,  
are used to study the ground-state properties of these systems, and
the results are found to be in excellent agreement 
both with each other and with results of exact 
calculations for the linear chain and results
of exact cumulant series expansions for lattices of 
higher spatial dimension.
The different techniques used  are compared and contrasted
in the light of these results, and the constructions of the 
approximate ground-state wave functions are especially discussed.
\end{abstract}

\pacs{PACS numbers: 75.10.Jm, 75.30.Gw, 75.50.Ee, 75.40.Cx}

\maketitle









\section{Introduction}

Two of the most versatile and most accurate semi-analytical formalisms 
of microscopic quantum many-body theory (QMBT) are the coupled cluster
method 
\cite{theory_ccm1,theory_ccm2,theory_ccm3,theory_ccm4,theory_ccm5,theory_ccm6,theory_ccm7,theory_ccm8} 
and the correlated basis function (CBF) method 
\cite{theory_cbf1,theory_cbf2,theory_cbf3,theory_cbf4,theory_cbf5,theory_cbf6,theory_cbf7,theory_cbf8,theory_cbf9,theory_cbf10,theory_cbf11}.
In recent years such QMBT methods, together with various quantum
Monte Carlo (QMC) techniques, have been applied with a great deal of
success to lattice quantum spin systems at zero temperature. Some typical
recent examples of such applications include Refs. 
\cite{ccm3,ccm8,ccm16,ccm18,ccm19,ccm20,ccm21,ccm22} 
for the
CCM, Refs. \cite{cbf1,cbf2,cbf3,cbf4} for the CBF method, and 
Refs. \cite{qmc1,qmc2,qmc3,qmc4,qmc5,qmc6,qmc7,qmc8} for the 
various QMC techniques. Current state of the art is such that these 
methods are sufficiently accurate to describe the various quantum
phase transitions between the states of different quantum order 
that exist in such abundance for spin-lattice systems. However, 
each of the above methods is characterised by its own strengths 
and weaknesses. Hence, a fuller and more complete understanding of 
such strongly interacting systems as the lattice quantum spin 
systems is expected to be given by the application of a range of 
such techniques than by the single application of any one of them. 
In this article we wish to apply the CCM, the CBF method, and the 
variational quantum Monte Carlo (VQMC) method to the spin-half 
transverse Ising model (for reviews of this model see, for example, Refs. 
\cite{new_ref1,new_ref2,new_ref3,new_ref4,new_ref5,exact_1D}). 
The Hamiltonian for this model on a lattice
of $N$ sites, each of which has $z$ nearest-neighbours, is given 
by
\begin{equation}
H = \biggl (\frac {z}2 + \lambda \biggr )N - \sum_{\langle i,j \rangle} 
\sigma_i^z \sigma_j^z - \lambda \sum_i \sigma_i^x ~~ ,
\label{H}
\end{equation}
where the $\sigma$-operators are the usual Pauli spin
operators and $\langle i,j \rangle$ indicates that each of the
$zN/2$ nearest-neighbour bonds on the lattice is counted once 
only. We work in the thermodynamic limit where $N \rightarrow
\infty$. We note that this model has an exact solution
in one dimension\cite{exact_1D}; and approximate techniques, 
such as the random phase approximation (RPA) \cite{rpa} and 
exact cumulant series expansions\cite{series1,series2}, 
have also been applied to it for lattices of higher spatial 
dimensionality. For $\lambda \ge 0$, we note furthermore 
that the model contains two distinct phases, with a critical
coupling strength $\lambda_c$ depending on lattice type and
dimensionality. For $\lambda < \lambda_c$ there is non-zero 
spin ordering in the $z$-direction, and hence this regime 
will be referred to here as the ferromagnetic regime. 
By contrast, for $\lambda > \lambda_c$, all of 
the ferromagnetic ordering is destroyed, and the classical
behaviour of these systems is that the spins lie along 
the positive $x$-axis. Hence, the $\lambda > \lambda_c$
regime will be referred to here as the paramagnetic regime. 
In Sec. II the technical aspects of applying the CCM, the 
CBF, and the VQMC methods to the spin-half transverse Ising
model are presented, and in Sec. III the results 
of these calculations are discussed. Finally, the 
conclusions are given in Sec. IV. 

\section{Quantum Many-Body Techniques}

\subsection{The Coupled Cluster Method (CCM)}
 
In this section, we firstly describe the general CCM formalism 
\cite{theory_ccm1,theory_ccm2,theory_ccm3,theory_ccm4,theory_ccm5,theory_ccm6,theory_ccm7,theory_ccm8},
and then proceed to apply it to the specific case of 
the spin-half transverse Ising model.
The exact ket and bra ground-state energy 
eigenvectors, $|\Psi\rangle$ and $\langle\tilde{\Psi}|$, of a 
many-body system described by a Hamiltonian $H$, 
\begin{equation} 
H |\Psi\rangle = E_g |\Psi\rangle
\;; 
\;\;\;  
\langle\tilde{\Psi}| H = E_g \langle\tilde{\Psi}| 
\;, 
\label{eq1} 
\end{equation} 
are parametrised within the single-reference CCM as follows:   
\begin{eqnarray} 
|\Psi\rangle = {\rm e}^S |\Phi\rangle \; &;&  
\;\;\; S=\sum_{I \neq 0} {\cal S}_I C_I^{+}  \nonumber \; , \\ 
\langle\tilde{\Psi}| = \langle\Phi| \tilde{S} {\rm e}^{-S} \; &;& 
\;\;\; \tilde{S} =1 + \sum_{I \neq 0} \tilde{{\cal S}}_I C_I^{-} \; .  
\label{eq2} 
\end{eqnarray} 
The single model or reference state $|\Phi\rangle$ is required to have the 
property of being a cyclic vector with respect to two well-defined Abelian 
subalgebras of {\it multi-configurational} creation operators $\{C_I^{+}\}$ 
and their Hermitian-adjoint destruction counterparts $\{ C_I^{-} \equiv 
(C_I^{+})^\dagger \}$. Thus, $|\Phi\rangle$ plays the role of a vacuum 
state with respect to a suitable set of (mutually commuting) many-body 
creation operators $\{C_I^{+}\}$, 
\begin{equation} 
C_I^{-} |\Phi\rangle = 0 \;\; , \;\;\; I \neq 0 \; , 
\label{eq3}
\end{equation} 
with $C_0^{-} \equiv 1$, the identity operator. These operators are 
complete in the many-body Hilbert (or Fock) space,  
\begin{equation} 
1=|\Phi\rangle \langle\Phi| + \sum_{I\neq 0} 
C_I^{+}  |\Phi\rangle \langle\Phi| C_I^{-} \; . 
\label{eq4}
\end{equation} 
Also, the {\it correlation operator} $S$ is decomposed entirely in terms 
of these creation operators $\{C_I^{+}\}$, which, when acting on the 
model state ($\{C_I^{+}|\Phi\rangle \}$), create excitations from it.
We note that although the manifest Hermiticity, 
($\langle \tilde{\Psi}|^\dagger = |\Psi\rangle/\langle\Psi|\Psi\rangle$), 
is lost, the intermediate normalisation condition 
$ \langle \tilde{\Psi} | \Psi\rangle
= \langle \Phi | \Psi\rangle 
= \langle \Phi | \Phi \rangle \equiv 1$ is explicitly 
imposed. The {\it correlation coefficients} $\{ {\cal S}_I, \tilde{{\cal S}}_I \}$ 
are regarded as being independent variables, even though formally 
we have the relation, 
\begin{equation} 
\langle \Phi| \tilde{S} =
\frac{ \langle\Phi| {\rm e}^{S^{\dagger}} {\rm e}^S } 
     { \langle\Phi| {\rm e}^{S^{\dagger}} {\rm e}^S |\Phi\rangle } \; . 
\label{eq5}
\end{equation} 
The full set $\{ {\cal S}_I, \tilde{{\cal S}}_I \}$ thus provides a complete 
description of the ground state. For instance, an arbitrary 
operator $A$ will have a ground-state expectation value given as, 
\begin{equation} 
\bar{A}
\equiv \langle\tilde{\Psi}\vert A \vert\Psi\rangle
=\langle\Phi | \tilde{S} {\rm e}^{-S} A {\rm e}^S | \Phi\rangle
=\bar{A}\left( \{ {\cal S}_I,\tilde{{\cal S}}_I \} \right) 
\; .
\label{eq6}
\end{equation} 

We note that the exponentiated form of the ground-state CCM 
parametrisation of Eq. (\ref{eq2}) ensures the correct counting of 
the {\it independent} and excited correlated 
many-body clusters with respect to $|\Phi\rangle$ which are present 
in the exact ground state $|\Psi\rangle$. It also ensures the 
exact incorporation of the Goldstone linked-cluster theorem, 
which itself guarantees the size-extensivity of all relevant 
extensive physical quantities. 

The determination of the correlation coefficients $\{ {\cal S}_I, \tilde{{\cal S}}_I \}$ 
is achieved by taking appropriate projections onto the ground-state 
Schr\"odinger equations of Eq. (\ref{eq1}). Equivalently, they may be 
determined variationally by requiring the ground-state energy expectation 
functional $\bar{H} ( \{ {\cal S}_I, \tilde{{\cal S}}_I\} )$, defined as in Eq. (\ref{eq6}), 
to be stationary with respect to variations in each of the (independent) 
variables of the full set. We thereby easily derive the following coupled 
set of equations, 
\begin{eqnarray} 
\delta{\bar{H}} / \delta{\tilde{{\cal S}}_I} =0 & \Rightarrow &   
\langle\Phi|C_I^{-} {\rm e}^{-S} H {\rm e}^S|\Phi\rangle = 0 ,  \;\; I \neq 0 
\;\; ; \label{eq7} \\ 
\delta{\bar{H}} / \delta{{\cal S}_I} =0 & \Rightarrow & 
\langle\Phi|\tilde{S} {\rm e}^{-S} [H,C_I^{+}] {\rm e}^S|\Phi\rangle 
= 0 , \;\; I \neq 0 \;\; . \label{eq8}
\end{eqnarray}  
Equation (\ref{eq7}) also shows that the ground-state energy at the stationary 
point has the simple form 
\begin{equation} 
E_g = E_g ( \{{\cal S}_I\} ) = \langle\Phi| {\rm e}^{-S} H {\rm e}^S|\Phi\rangle
\;\; . 
\label{eq9}
\end{equation}  
It is important to realize that this (bi-)variational formulation 
does {\it not} lead to an upper bound for $E_g$ when the summations for 
$S$ and $\tilde{S}$ in Eq. (\ref{eq2}) are truncated, due to the lack of 
exact Hermiticity when such approximations are made. However, it is clear 
that the important Hellmann-Feynman theorem {\it is} preserved in all 
such approximations. 

We also note that Eq. (\ref{eq7}) represents a coupled set of 
nonlinear multinomial 
equations for the {\it c}-number correlation coefficients $\{ {\cal S}_I \}$. 
The nested commutator expansion of the similarity-transformed Hamiltonian,  
\begin{equation}  
\hat H \equiv {\rm e}^{-S} H {\rm e}^{S} = H 
+ [H,S] + {1\over2!} [[H,S],S] + \cdots 
\;\; , 
\label{eq10}
\end{equation} 
together with the fact that all of the individual components of 
$S$ in the sum in Eq. (\ref{eq2}) commute with one another, imply 
that each element of $S$ in Eq. (\ref{eq2}) is linked directly to
the Hamiltonian in each of the terms in Eq. (\ref{eq10}). Thus,
each of the coupled equations (\ref{eq7}) is of linked cluster type.
Furthermore, each of these equations is of finite length when expanded, 
since the otherwise infinite series of Eq. (\ref{eq10}) will always 
terminate at a
finite order, provided (as is usually the case) that each term in the 
second-quantised form of the Hamiltonian $H$ contains a finite number of 
single-body destruction operators, defined with respect to the reference 
(vacuum) state $|\Phi\rangle$. Therefore, the CCM parametrisation naturally 
leads to a workable scheme which can be efficiently implemented 
computationally. It is also important to note that at the heart
of the CCM lies a similarity transformation, in contrast with  
the unitary transformation in a standard variational formulation 
in which the bra state $\langle\tilde\Psi|$ is simply taken as
the explicit Hermitian adjoint of $|\Psi\rangle$. 

The CCM formalism is exact in the limit of inclusion of
all possible multi-spin cluster correlations for 
$S$ and $\tilde S$, although in any real application 
this is usually impossible to achieve. It is therefore 
necessary to utilise various approximation schemes 
within $S$ and $\tilde{S}$. The three most commonly 
employed schemes previously utilised have been: 
(1) the SUB$n$ scheme, in which all correlations 
involving only $n$ or fewer spins are retained, but no
further restriction is made concerning their spatial 
separation on the lattice; (2) the SUB$n$-$m$  
sub-approximation, in which all SUB$n$ correlations 
spanning a range of no more than $m$ adjacent lattice 
sites are retained; and (3) the localised LSUB$m$ scheme, 
in which all multi-spin correlations over distinct 
locales on the lattice defined by $m$ or fewer contiguous 
sites are retained. The specific application of the CCM 
to the spin-half transverse Ising model in the paramagnetic and ferromagnetic 
regimes is now described.

\subsubsection{The Paramagnetic Regime}

In the paramagnetic regime, a model state is utilised in 
which all spins point along the $x$-axis, although it is 
found to be useful to rotate the local spin coordinates 
of these spins such that all spins in the model state point 
in the `downwards' direction (i.e., along the negative
$z$-axis). This (canonical) transformation is given by,
\begin{equation}
\sigma^x \rightarrow -\sigma^z ~~,~~ 
\sigma^y \rightarrow  \sigma^y ~~,~~ 
\sigma^z \rightarrow  \sigma^x ~~,
\end{equation}
such that the transverse Ising Hamiltonian of Eq. (\ref{H}) is 
now given in the (rotated) spin-coordinate frame by,
\begin{equation}
  H  =  \biggl (\frac z2 + \lambda \biggr )N - \sum_{\langle i,j \rangle} \biggl [ 
  \sigma_i^+ \sigma_j^+ +  \sigma_i^- \sigma_j^- +
  \sigma_i^- \sigma_j^+ +  \sigma_i^+ \sigma_j^-  \biggr ]
  + \lambda \sum_i \sigma_i^z  ~~,
\end{equation}
where $\sigma_k^{\pm} \equiv \frac 12 (\sigma_k^x \pm {\rm i} \sigma_k^y)$. 
In these local coordinates the model state is thus the ``ferromagnetic''
state $| \Psi \rangle = | ~ \downarrow \downarrow \cdots \downarrow \cdots 
\rangle$ in which all spins point in the downwards direction.
In order to reflect the symmetries of this Hamiltonian, the cluster 
correlations within $S$ are explicitly restricted to those for which 
$s_T^z = \sum_i s_i^z$ (in the rotated coordinate frame) is an even 
number. Hence, the LSUB2 approximation is defined by
\begin{equation}
S=\frac {b_1}2 ~ \sum_{i,\rho} \sigma_i^+ \sigma_{i+\rho}^+ ~~ ,
\end{equation}
where $\rho$ covers all nearest-neighbour lattice vectors.
The ground-state energy is now given in terms of $b_1$ by,
\begin{equation}
\frac {E} N   = \frac {z}{2} (1-b_1) ~~ .
\label{eq11}
\end{equation}
It is found that this expression is
valid for any level of approximation for $S$. Using 
Eq. (\ref{eq6}) it is found that,
\begin{equation}
5b_1^2  + 4\lambda b_1 - 1  =  0 ~~ ,
\label{eq_lsub2para}
\end{equation}
and hence an approximate solution for the ground-state 
energy at the LSUB2 approximation level purely in terms of 
$\lambda$ may be obtained. The SUB2 approximation contains 
all possible two-body correlations, for a given lattice,
and is defined by
\begin{equation}
S=\frac 12 \sum_{i} \sum_r b_r ~ \sigma_i^+ \sigma_{i+r}^+ ~~ ,
\end{equation}
where the index $r$ indicates a lattice vector.
Eq. (\ref{eq6}) may once again be utilised to determine
the SUB2 ket-state equations. Hence the CCM SUB2 ket-state 
equation corresponding to a two-body correlation characterised 
by index $s$ is given by,
\begin{eqnarray}
(1+2b_1^2)\delta_{s,\rho} - 4b_1 b_{s} + \sum_{r} b_{r} b_{r+ s
+ \rho} + 2b_{s+ \rho} -4 \lambda b_{s} &=& 0 ~~ .
\label{sub2}
\end{eqnarray}
We note that this equation is meaningful only for $s \ne 0$ 
as we may only ever have one Pauli raising operator per
lattice site. This equation may be solved by performing 
a Fourier transformation. (Details of how this is
achieved in practice are not given here and the interested 
reader is referred to Refs. \cite{ccm3,ccm18}.) 
An alternative approach, however, is to use Eq. (\ref{sub2}) 
in order to fully define the SUB2-$m$ equations. 
This is achieved by truncating the range of the 
two-body correlations (i.e., by setting $|s| 
\le m$), and the corresponding SUB2-$m$ equations 
may be solved numerically via the Newton-Raphson 
technique (or other such techniques). We note that
coupled sets of high-order LSUB$m$ equations
may be derived using computer-algebraic 
techniques, as discussed in Ref. \cite{ccm19}.
The technicalities of these calculations are not
considered here, but the interested reader is referred
to Ref. \cite{ccm19}. A full discussion of the CCM 
results based on the paramagnetic model state is 
deferred until Sec. III.

\subsubsection{The Ferromagnetic Regime}

In the ferromagnetic regime, a model state is chosen in which 
all spins point `downwards' (along the negative $z$-axis), 
and so the Hamiltonian of Eq. (\ref{H}) may therefore be utilised 
directly within the CCM calculations. The lowest order approximation 
is the now SUB1 approximation (in which case, $S = a \sum_i \sigma_i^+$) 
and the ground-state energy is given in terms of $a$ by,
\begin{equation}
\frac {E} N  = \lambda (1 - a) ~~ .
\label{eq12}
\end{equation}
It is again noted that this expression is valid for 
any level of approximation in $S$. In this case, it is 
found that the solution of the SUB2 approximation collapses onto 
the LSUB2 solution due to the simple nature of the Hamiltonian 
and model state, although it is again possible to perform 
high-order LSUB$m$ calculations. Furthermore, the lattice magnetisation 
(i.e., the magnetisation in the $z$-direction), $M$, is defined 
within the CCM framework by,
\begin{equation}
M = - \frac 1N \sum_{i=1}^N  \langle \tilde 
\Psi \mid \sigma_i^z \mid \Psi \rangle ~~ ,
\label{eq13}
\end{equation}
which may be determined once both the ket- and bra-state
equations have been solved at a given level of approximation. 
Again, the discussion of the results for this model state is
deferred until Section III.

\subsection{The CBF Formalism}

The treatment of the transverse Ising model by the CBF 
method is begun by defining the lattice magnetisation 
(i.e., again the magnetisation in the $z$-direction), given by
\begin{equation}
M = \frac {\langle \psi \mid \sigma_i^z \mid \psi \rangle}
{\langle \psi | \psi \rangle} ~~,
\label{eq14}
\end{equation}
for a ground-state trial wave function, $| \psi \rangle$.
Furthermore, the `transverse' magnetisation is given by,
\begin{equation}
A = \frac {\langle \psi \mid \sigma_i^x \mid \psi \rangle}
{\langle \psi | \psi \rangle} ~~.
\label{eq15}
\end{equation}
It is also found to be useful to define a spatial 
distribution function (which plays a crucial
part in any CBF calculation) in the following 
manner,
\begin{equation}
g({\bf n}) = \frac {\langle \psi \mid \sigma_i^z \sigma_j^z  \mid \psi \rangle}
{\langle \psi | \psi \rangle} ~~ ,
\label{eq16}
\end{equation}
where ${\bf n}={\bf r}_i - {\bf r}_j$. The corresponding approximation
to the ground-state energy per spin is given by
\begin{equation}
  \frac {E}N = \frac {\langle \psi | H | \psi \rangle}  
  {N \langle \psi | \psi \rangle} = \frac z2 
  - \sum_{{\bf n}} \Delta({\bf n}) 
  g({\bf n}) +  \lambda(1-A) ~~,
  \label{eq17}
\end{equation}
where the function $\Delta({\bf n})$ is equal to unity when ${\bf n}$ 
is a nearest-neighbour lattice vector and is zero elsewhere.
It is noted that the distribution function $g({\bf n})$ may be decomposed 
according to $g({\bf n})=\delta_{{{\bf n}},0} + (1-\delta_{{{\bf n}},0})M^2
+ (1-M^2)G({\bf n})$ such that $G({\bf n})$ now contains the 
short-range part of the spatial distribution function
and vanishes in the limit $|{\bf n}| \rightarrow \infty$.
The magnetisation, $M$, and the transverse magnetisation, $A$, 
may now be expressed in a factorised form in terms of a 
`spin-exchange strength', $n_{12}$, such that, 
\begin{equation}
A = (1-M^2)^{\frac 12} n_{12} ~~.
\label{eq18}
\end{equation}
The energy functional is now expressed in terms of $G({\bf n})$ 
and $n_{12}$ as,
\begin{equation}
\frac {E}N = (1-M^2) \biggl \{ \frac z2 - \sum_{{\bf n}}
\Delta({\bf n}) G({\bf n}) \biggr \}
+  \lambda \biggl \{ 1-(1-M^2)^{\frac12}n_{12} \biggr \} ~~.
\label{eq19}
\end{equation}
Note that in the mean-field approximation $G({\bf n})$ in Eq. 
(\ref{eq19}) is set to zero (for all ${\bf n}$) and $n_{12}$ is set 
to unity. 

In order to determine the ground-state energy and 
other such ground-state expectation values, a Hartree-Jastrow 
Ansatz is now introduced, given by
\begin{equation}
| \psi \rangle = {\rm exp} \{ MU_M + U \} |0\rangle ~~ .
\label{eq20}
\end{equation}
The reference state $|0\rangle$ is a tensor product of spin states
which have eigenvalues of $+1$ with respect to $\sigma^x$. 
The correlation operators $U$ and $U_M$ are written in 
terms of pseudopotentials, $u({\bf r}_{ij})$, $u_1({\bf r}_{i})$, and
$u_M({\bf r}_{ij})$, where
\begin{equation}
U = \frac 12 \sum_{i<j}^N u({\bf r}_{ij}) \sigma_i^z
\sigma_j^z ~~ ,
\label{eq21}
\end{equation}
and
\begin{equation}
U_M = \sum_i^N u_1({\bf r}_i) \sigma_i^z + \frac 14
\sum_{i<j}^N u_M({\bf r}_{ij}) (\sigma_i^z + 
\sigma_j^z) ~~ .
\label{eq22}
\end{equation}
The pseudopotential $u_1({\bf r}_i)\equiv u_1$ is independent of the lattice
position by translational invariance, and the pseudopotentials, 
$u({\bf r}_{ij})$ and $u_M({\bf r}_{ij})$, similarly depend only on the 
relative distance, 
$|{\bf n}| = |{\bf r}_i - {\bf r}_j| \equiv |{\bf r}_{ij}|$. 
The Jastrow correlations are determined via a cluster 
expansion of the various quantities in the Hamiltonian,
as explained in Refs. \cite{cbf1,cbf2,cbf3,cbf4}. A common 
approximation yields the hypernetted chain (HNC/0) 
equations, which one may solve iteratively in order to determine
the Hartree-Jastrow pseudopotentials. One then wishes
to determine the expectation values such as the ground-state 
energy, and in the paramagnetic regime an explicit assumption
is made that $M=0$. However, in the ferromagnetic regime
$M$ is taken to be a variational parameter with respect
to the ground-state energy of Eq. (\ref{eq19}).

There are now two ways of determining the pseudopotentials
from the HNC equations. The first such approach is to assume 
that the pseudopotential has the simple parametrised form,
\begin{equation}
u({\bf n}) = \alpha ~ \Delta({\bf n})~~ ,
\label{eq23}
\end{equation}
where $\Delta({\bf n})$ is unity if ${\bf n}$ is a nearest-neighbour 
lattice vector and is zero otherwise. This approach is henceforth
denoted as the parametrised HNC CBF method. The $c$-number $\alpha$ is 
taken to be a variational parameter with respect to which the
ground-state energy is minimized. In the paramagnetic regime, the minimum 
of the energy surface as a function of $\alpha$ is sought, 
at a given value of $\lambda$. This is easily performed 
computationally, and the solution is readily tracked iteratively,
starting from the trivial limit $\lambda \rightarrow \infty$
and then moving to smaller values of $\lambda$. In the 
ferromagnetic regime, one again searches 
for a minimum of the energy surface, but this time with 
respect to both $\alpha$ and $M$, at a given value of 
$\lambda$. In this case, one tracks from the trivial limit of 
$\lambda=0$ to higher values of $\lambda$.
In previous articles\cite{cbf1,cbf2,cbf3,cbf4}, this was 
achieved by analytically determining the derivative of the 
energy with respect to $M$, although in this article   
a computational minimisation of the energy is performed 
with respect to both variables. The second such approach 
allows one to find the optimal pseudopotential 
within the CBF/HNC framework from a functional 
minimisation,
\begin{equation}
\frac {\delta E } {\delta u({\bf n})} = 0 ~~ .
\label{eq24}
\end{equation}
Within the context of this article, this is henceforth
denoted as the {\it paired phonon approximation} 
(PPA) (or, more precisely, the paired magnon approximation),
and the corresponding equations are the PPA equations. 
Note that in this article the PPA calculation is only 
performed in the paramagnetic regime, although PPA 
results in the ferromagnetic regime have also been 
performed previously\cite{cbf2,cbf3}. (Indeed, results for the 
phase transition points predicted by the PPA CBF approach 
of Ref.\cite{cbf2,cbf3} are quoted in the Table \ref{tab1}.) 
A full discussion of the results of the CBF calculations 
presented in this section, in comparison to the corresponding results 
of the CCM and the VQMC method, is given in Section III.

\subsection{The VQMC Formalism}

Although the specific variational calculations presented 
in this article concentrate on the 
spin-half transverse Ising model, we note that the 
formalism presented in this section is given in a 
generalised form and that the treatment of other spin 
models would follow a similar pattern. We 
shall specifically consider here the transverse Ising 
model in the ferromagnetic regime where the relevant
Hamiltonian is defined by Eq. (\ref{H}). An Ansatz 
for the expansion coefficients, $\{ c_I \}$, of a
ground-state wave function defined by
\begin{equation}
  |\psi\rangle = \sum_I c_I |I\rangle ~~,
  \label{eq25}
\end{equation}
is chosen. Note that $\{ |I\rangle \}$ denotes a 
complete set of Ising basis states, defined as all possible
tensor products of states on all sites having eigenvalues
$\pm 1$ with respect to $\sigma^z$. An expression for 
the ground-state energy is thus given by, 
\begin{equation}
  E = \frac {\sum_{I_1,I_2} c_{I_1}^* c_{I_2} 
  \langle I_1 | H | I_2 \rangle}
  {\sum_{I'} |c_{I'}|^2} ~~ .
  \label{eq26}
\end{equation}
Specifically for the spin-half transverse Ising model, a Hartree-Jastrow 
Ansatz \cite{variational_1D} (for $\lambda >0$) is now defined with respect to 
the $\{ c_I \}$ expansion coefficients, where
\begin{equation}
c_I = \langle I | \prod_{k} \biggl ( 1 + a_k P_{k}^{\uparrow} 
\biggr ) \prod_{i<j} \biggl ( 1 + f_{i,j} \biggl [ P_{i}^{\uparrow} 
P_{j}^{\downarrow} + P_{i}^{\downarrow} P_{j}^{\uparrow} 
\biggr ] \biggr ) |I\rangle ~~ .
\label{eq27}
\end{equation}
The $P^{\uparrow}$ and $P^{\downarrow}$ are the usual 
projection operators of the spin-half `up' and `down' 
states respectively. 
The simplest form of the variational Ansatz of Eq. 
(\ref{eq27}) is given by, 
\begin{equation}
f_{i,j} = 
~~~\left \{~~~
\mbox{
\begin{tabular}{ll}
$f_1$ &  if $i$ and $j$ are nearest neighbours, \\
0     &  otherwise. 
\end{tabular}
}
\right .
\label{eq28}
\end{equation}
The symmetry-breaking term $a_k (= a)$ is also
independent of $i$ by translational invariance. 
The expectation value of Eq. (\ref{eq26}) 
may now be evaluated directly and the variational 
ground-state energy minimised with respect to 
both $a$ and $f_1$ at each value of $\lambda$.
However, such a calculation is soon limited by 
the rapidly increasing set of Ising states and 
the amount of computational power available. 
Indeed, for the spin-half transverse Ising model 
the number of states that one must sum over is
$2^N$, where $N$ is the number of lattice sites. 
For the linear chain it is possible to solve for 
chains of length $N \stackrel{<}{\sim} 12$ with relatively little
computational difficulty, although the calculations
with $N >12$ grow rapidly in computational difficulty.

Hence, as an alternative for lattices of larger
size, we may simulate the summation over all the states 
$I_1$ in Eq. (\ref{eq26}). In order to to do this we define 
the probability distribution for the set of states 
$\{ |I\rangle \}$, given by
\begin{equation}
P(I)  \equiv \frac {|c_{I}|^2}{\sum_{I'} |c_{I'}|^2} ~~ ,
\label{eq29}
\end{equation}
and the {\it local energy} of these states, given by
\begin{equation}
E_L(I) \equiv \sum_{I_1} \frac {c_{I_1}}{c_{I}} 
\langle I | H | I_1 \rangle ~~ .
\label{eq30}
\end{equation}
The expression of Eq. (\ref{eq26}) may thus be equivalently 
written as,
\begin{equation}
E  = \sum_{I} ~ P(I) ~ E_L(I) ~~ .
\label{eq31}
\end{equation}
We now wish to perform a random walk based on the probability
distribution of Eq. (\ref{eq29}). However, a few more 
useful quantities are best defined before a detailed 
description of the VQMC algorithm is actually given.
Firstly, the acceptance probability, $A(I \rightarrow I')$, 
of Monte Carlo `move' from  state $|I\rangle$ to state 
$|I'\rangle$ is given by
\begin{equation}
A(I \rightarrow I') = {\rm min} ~ [1,q(I \rightarrow I')] ~~ ,
\label{eq32}
\end{equation} 
where 
\begin{equation}
q(I \rightarrow I') \equiv \frac {P(I') T(I' \rightarrow I)} 
{P(I) T(I \rightarrow I')} ~~ .
\label{eq33}
\end{equation} 
Here, $T(I \rightarrow I')$ is the sampling distribution function. For
spin lattice problems, if state $|I\rangle$ can connect to $K(I)$ 
possible Ising states via the off-diagonal elements of the 
Hamiltonian then $T(I \rightarrow I')=1/K(I)$. Hence, 
$q(I \rightarrow I')$ is written as
\begin{equation}
q(I \rightarrow I') = \frac {P(I') K(I)}{P(I) K(I')} ~~ .
\label{eq34}
\end{equation} 
For the spin-half transverse Ising model, we note that 
$K(I)$ is equal to $N$ for any state $|I\rangle$ and so the common
factor of $N$ in Eq. (\ref{eq34}) cancels. The simplest
VQMC procedure is now defined by the following algorithm:

\begin{enumerate}

\item Select an initial Ising state $|I\rangle$ for which $\langle I|
  \Psi \rangle \ne 0$, where $| \Psi \rangle$ is the `true' ground-state wave 
  function of the system.

\item Choose a particular state $|I'\rangle$ out of the $K(I)$ possible 
  states accessible to $|I\rangle$ via the off-diagonal elements in 
  $H$. 

\item Define a random number $r$ in the range $[0,1]$ and accept 
  this move from state $|I\rangle$ to state $|I'\rangle$ if and only
if

\begin{equation}
A(I \rightarrow I') > r ~~ .
\label{eq35}
\end{equation}

\item If the move is accepted then let $I \rightarrow I'$ and
  $c_{I} \rightarrow c_{I'}$.

\item Obtain the local energy $E(I)$ of Eq. (\ref{eq30}) for state 
  $|I\rangle$.

\item Repeat from stage (2) $N_{MC}$ times.

\item The average ground-state energy (and the error therein) 
  may be determined from the $N_{MC}$ number of local energies
  during the simulation.

\end{enumerate}
The minimal VQMC ground-state energy is now obtained 
by searching over the variational parameter space for
either the lowest ground-state energy or lowest variance
in the ground-state energy (here for a given value of
$\lambda$ in  $H$). 
In order to determine the lattice magnetisation, we note 
that
\begin{eqnarray}
M & = & \biggl | \frac {\langle \psi | \sum_i \sigma_i^z | \psi \rangle}
{N \langle \psi | \psi \rangle} \biggr | ~~, \nonumber \\
  & = & \biggl | \sum_I P(I) M_L(I) \biggr | ~~, 
\end{eqnarray}
where $P(I)$ is the probability density function given above, and
\begin{equation}
 M_L(I) \equiv \sum_{I_1} \frac {c_{I_1}}{N c_{I}} 
\langle I | \sum_i \sigma_i^z | I_1 \rangle =
\frac 1N  \langle I | \sum_i \sigma_i^z | I \rangle ~~ .
\end{equation}
Hence, a mean value (and its associated error) for the VQMC 
lattice magnetisation, $M$, may be obtained by determining 
the average of the local lattice magnetisation, $M_L(I)$, 
throughout the lifetime of the run. A discussion of the 
results of the variational calculations discussed here 
is given in Section III.

\subsection{The Infinite Lattice Limit and Convergence of Results}

In this section we consider how the results of each method are 
determined in the infinite lattice limit. Firstly, it is noted that the 
CCM method produces expectation values which are size-extensive 
(i.e., the numerical values of each expectation value scale 
linearly with $N$), and we always deal with an infinite
lattice in all calculations from the very outset. Furthermore, 
the `raw' CCM LSUB$m$ results based on the ferromagnetic 
model state are found to converge rapidly with increasing 
LSUB$m$ approximation level $m$ over most of the 
ferromagnetic regime except for a region very near 
to the phase transition point. In order 
to obtain even better results for the CCM method across the 
whole of this regime, a simple extrapolation of the LSUB$m$ 
data in the limit $m \rightarrow \infty$ has also been 
carried out at each value of $\lambda$ separately. The 
results of the extrapolation using a leading-order 
`power-law' dependence (see Appendix A for details) are 
denoted as Extrapolated(1) CCM results and results of a Pad\'e 
approximation for $l=0$ (again, see Appendix A for details) 
are called Extrapolated(2) CCM results. In the paramagnetic
regime, the results for the ground-state energy are found
to converge extremely rapidly with LSUB$m$ approximation
level over the whole of this regime and so no extrapolation 
of these results is necessary.

For the CBF method, although the treatment presented here is 
formally valid for any lattice size (including the infinite 
lattice case), the results presented have been obtained for finite-sized 
lattices. The results are found to converge extremely 
rapidly with increasing lattice size, and the results of 
the 20$\times$20 square lattice (used in the figures 
given below) are found to be essentially fully converged
for all $\lambda$ except for a region very near to the critical 
point. 

The results for the VQMC method presented below have
been obtained for a 16$\times$16 square lattice, 
where the number of Monte Carlo iterations was set to 50000. 
As for the CBF calculations, the 16$\times$16 lattice is 
again expected to be large enough for the VQMC results to be essentially 
fully converged to the infinite lattice limit for all values of
$\lambda$ except for a region near to the critical point. 
By comparing the results of a 10$\times$10 lattice VQMC 
calculation with those of a 16$\times$16 lattice it was found
that this was indeed true. 
Furthermore, we note that for small $\lambda$ the variational 
minimum of the ground-state energy was found to be rather 
flat with respect to the variational parameters $a$ and 
$f_1$, and the ground-state energy was also highly converged 
with increasing lattice size. However, as the phase transition 
point is approached one finds that a precise evaluation 
of the position of the variational minimum with respect to 
$a$ and $f_1$ becomes harder to determine. 

\section{Results}

\begin{figure}
\epsfxsize=10cm
\centerline{\epsffile{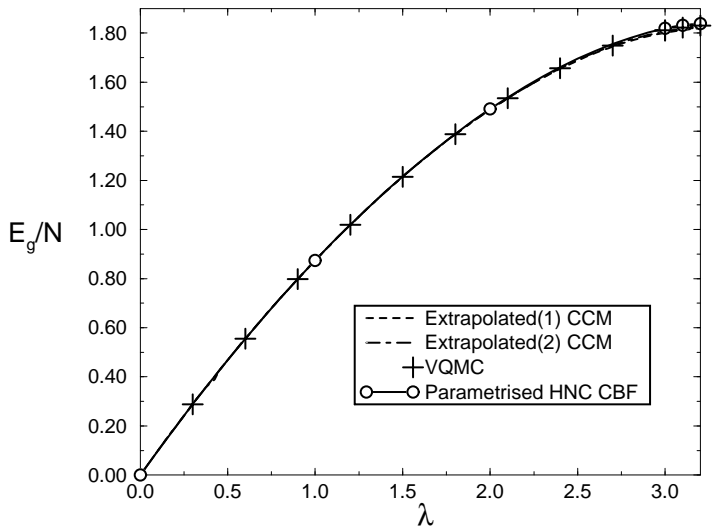}}
\caption{Results for the ground-state energy per spin, $E_g/N$,  
of the spin-half
transverse Ising model on the square lattice in the ferromagnetic 
regime using the CCM, CBF, and VQMC approaches.} 
\label{fig1}
\end{figure}

The results for the ground-state energy per spin of the spin-half 
transverse Ising model on the square lattice in the ferromagnetic 
regime are shown in Fig. \ref{fig1}. We see from 
this figure that excellent correspondence between the results 
of the CCM, CBF, and VQMC methods for the ground-state energy 
is obtained in this regime. We note, however, that the extrapolated
CCM results appear to lie very slightly lower than the other two
sets of results, especially near to the phase transition point.
This indicates the increasing importance of higher-order correlations
for the ground-state energy near to the phase transition point. 

The results for the lattice magnetisation $M$ obtained 
using the CCM, CBF, and VQMC formalisms are shown in 
Fig. \ref{fig2} (and also Fig. \ref{fig4} in Appendix A) 
for the spin-half transverse Ising model on the square 
lattice. We note that the `raw' CCM 
LSUB$m$ results for the lattice magnetisation do not 
become zero at any value of $\lambda$ for any finite 
value of the truncation index $m$, because all of the 
ferromagnetic order inherent in the model state must be 
destroyed in order for $M$ to be zero in this case. In 
practice this is a difficult thing for the CCM to achieve 
with this model state. However, we may see from Fig. \ref{fig2} 
that the extrapolated CCM results are in excellent 
agreement with those results of the CBF and VQMC methods.
In addition, it is possible to imagine other CCM model states, 
such as a spin-flop model state or a mean-field model 
state, in which the lattice magnetisation with respect 
to this state is not {\it a priori} fully saturated for 
all $\lambda$. Furthermore, we remark that a treatment 
of this problem with the ferromagnetic model state using 
the extended coupled cluster method (ECCM) 
\cite{theory_ccm4,theory_ccm5,theory_ccm7,ccm23}
would present an interesting challenge. 

\begin{figure}
\epsfxsize=10cm
\centerline{\epsffile{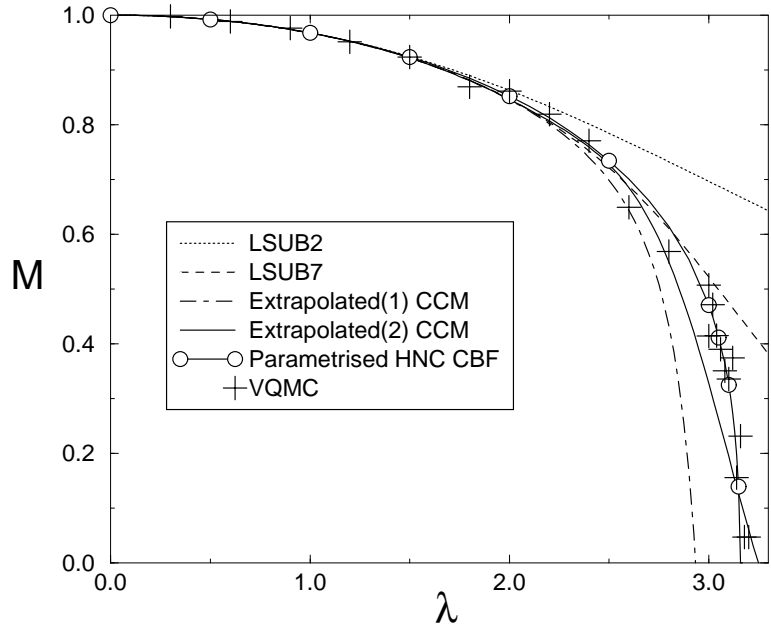}}
\caption{Results for the lattice magnetisation, $M$, 
of the spin-half transverse Ising model on the square 
lattice using the CCM, CBF, and VQMC approaches.} 
\label{fig2}
\end{figure}

\begin{figure}
\epsfxsize=10cm
\centerline{\epsffile{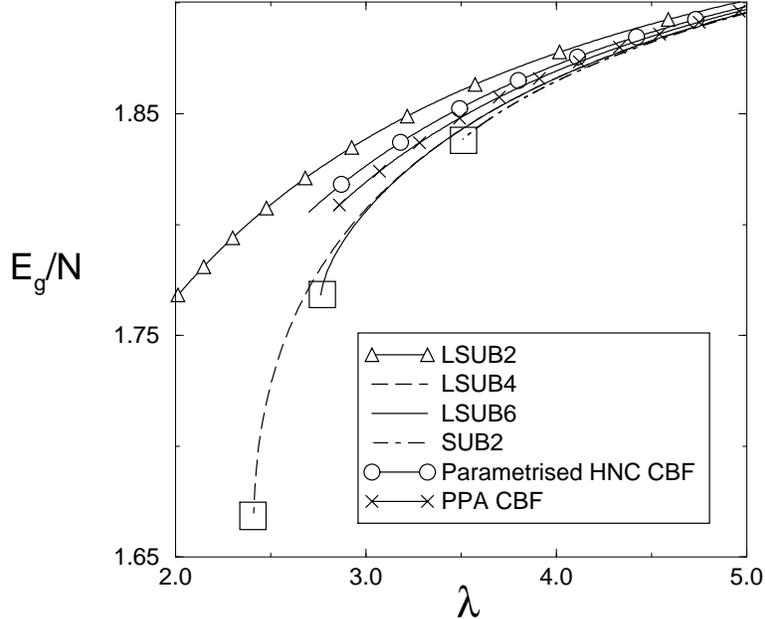}}
\caption{Results for the ground-state energy per spin,
$E_g/N$, of the spin-half transverse Ising model on 
the square lattice in the paramagnetic regime using 
the CCM and CBF approaches. The boxes for the CCM 
data indicate the terminating points at which $\chi$
becomes infinite.} 
\label{fig3}
\end{figure}

In the paramagnetic regime, the ground-state energy per spin
for the transverse Ising model on the square lattice
is presented in Fig. \ref{fig3}. The `raw' CCM LSUB$m$ 
results for the ground-state energy are already highly 
converged with increasing truncation index $m$, even up 
to the phase transition point. We note again that an 
extrapolation in the limit $m \rightarrow \infty$ is 
therefore not necessary. Indeed, good correspondence 
between the results of the different methods is seen 
although it is noted that CCM LSUB4 and LSUB6 ground-state 
energies lie lower than those predicted by the CBF. 
This indicates that high-order order correlations
become increasingly important the nearer one gets
to the phase transition point. However, this could be 
rectified, in principle, for the CBF method by the 
inclusion of higher-order (than pairwise Jastrow) 
correlations in the ground-state wave function. We
note that in practice, however, the inclusion of 
such higher-order correlations in the ground-state wave 
function in the CBF and VQMC methods is a difficult and 
unresolved question.

It is also possible to determine the second-derivative
of the ground-state energy per spin with respect to $\lambda$
for the CCM calculations based on the paramagnetic model state,
defined by
\begin{equation}
\chi \equiv - \frac 1N \frac {\partial^2 E_g} {\partial \lambda^2} ~.
\end{equation}
It is found that $\chi$ diverges at some critical 
value $\lambda_c$ for the SUB2 approximation in any 
dimension, and for SUB2-$m$ and LSUB$m$ (with $m \ge 4$)
approximations for spatial dimensionality greater 
than one. Again, this behaviour is associated with 
a phase transition in the real system and the point 
at which this occurs is denoted $\lambda_c$. 
Correspondingly, CCM results for $\lambda < \lambda_c$
based on the paramagnetic model state do no exist,
and hence $\lambda_c$ acts as a terminating point for the
calculation in the paramagnetic regime. Also, it is 
found that the SUB2-$m$ results for $\lambda_c$ as a 
function of $m$ scale with $m^{-2}$ and a simple linear 
extrapolation gives the full SUB2 result for the critical 
point to within a $2\%$ accuracy. By analogy, this 
rule has also been used for the LSUB$m$ results to 
extrapolate to the limit $m \rightarrow \infty$, and 
the results thus determined are shown in Table \ref{tab1}. 
The values thus obtained are also in good agreement
with the points at which $M \rightarrow 0$ from 
the extrapolations in the ferromagnetic regime discussed 
above (and see Fig. 2 for the square lattice case).

For the CBF and VQMC methods the point, in terms of 
$\lambda$, at which $M$ becomes zero is taken to 
indicate a quantum phase transition and is again
denoted, $\lambda_c$, and these results are presented
in Table \ref{tab1}. The phase transition point predicted
by the VQMC method on the square lattice case is estimated 
to be at $\lambda_c=3.15\pm0.05$. For the linear chain, the
expression for ground-state energy of Eq. (\ref{eq26}) 
has been obtained directly for chains with $N \le 12$. 
These results are found to be in good agreement with 
a previous calculation\cite{variational_1D} using the 
Ansatz of Eq. (\ref{eq27}) for the linear chain 
transverse Ising model which predicted a value for 
the phase transition point of $\lambda_c=1.206$. 


\bigskip

\begin{table}[ht!]
\caption{Results for the critical points of the spin-half 
transverse Ising model on various lattices using the CCM 
and CBF approaches. These results are compared to those 
of exact calculations for the linear chain \cite{exact_1D} and to 
RPA \cite{rpa} and exact cumulant series expansions \cite{series2} in 
higher spatial dimensionality. For the linear chain, we 
note that the LSUB$m$ approximation does not show any 
evidence of a critical point, at the levels of approximation 
shown, and this is indicated by `none'. (Previous CBF results of 
Refs. \cite{cbf1,cbf2,cbf3} for the square and cubic lattices are also 
appropriately indicated.)}
\begin{center}
\begin{tabular}{|l|c|c|c|c|}
 \toprule
                        &Linear Chain&Square        &Triangular     &Cubic     
\\ \hline\hline
Classical               &2          &4              &6              &6         
\\ \hline
RPA\footnote[1]{from Ref. \cite{rpa}}           
                        &--         &3.66           &--             &5.76      
\\ \hline
CCM $\lambda_c$ SUB2    &1.44       &3.51           &5.42           &5.55     
\\ \hline
CCM $\lambda_c$ LSUB4   &none       &2.41           &4.27           &3.85      
\\ \hline
CCM $\lambda_c$ LSUB6   &none       &2.76           &4.57           &4.61      
\\ \hline
CCM $\lambda_c$ LSUB$\infty$
                        &none       &3.04           &4.81           &5.22      
\\ \hline
parametrised HNC CBF    &1.22       &3.12\footnote[2]{from Ref. \cite{cbf1}}
                        &4.91       &5.17[{\it b]}  \\ \hline
PPA CBF\footnote[3]{from Refs. \cite{cbf2,cbf3}}    
                        &--          &3.14           &--             &5.10 
\\ \hline
Variational or VQMC Calculations     &1.206\footnote[4]{from Ref. \cite{variational_1D}}
                                    &3.15$\pm$0.05  &--             &--
\\ \hline
Exact or Series Expansions  Calculations 
                        &1.0\footnote[5]{from Ref. \cite{exact_1D}}
                        &3.044\footnote[6]{from Ref. \cite{series2}}
                        &4.768[{\it f]}
                        &5.153[{\it f]}\\ \hline
\end{tabular}
\end{center}
\label{tab1}
\end{table}

\section{Conclusions}

In this article, results of the CCM, CBF, 
and VQMC approaches for the ground-state 
energy, the lattice 
magnetisation, and the position of the critical 
point of the spin-half transverse Ising model 
on various lattices have been presented. These 
results have been seen to be in excellent agreement 
both with each other and with those results of exact 
calculations for the linear chain and those of
exact cumulant series expansions for higher 
spatial dimensionality. Indeed, by treating 
these systems using three separate approaches,
it has been shown that each set of results has been 
mutually supported and reinforced by those of 
the other approaches.

Furthermore, we have gained some insight into 
the strengths and weaknesses of each approach.
This is exemplified in the different parametrisations
of the ground-state wave function. The CBF 
and VQMC approaches both utilise 
Jastrow wave functions and their bra states 
are always the explicit Hermitian adjoints of the 
corresponding ket states. Hence, for the CBF and 
VQMC approaches, an upper bound to the 
true ground-state energy is, in principle, 
obtainable, although the approximations 
made in calculating the energy may destroy 
it. By contrast, the CCM uses a bi-variational 
approach in which the bra and ket states 
are not manifestly constrained to be 
Hermitian adjoints and hence an upper
bound to the true ground-state energy is
not necessarily obtained. Also, the CCM uses creation operators 
with respect to some suitably normalised 
model state in order to span the complete 
set of (here) Ising states. The other approaches,
in essence, use projection operators to 
form the Hartree and the Jastrow correlations.
For the CBF case, this is with respect to
a reference state, whereas for the 
VQMC case, the Hartree-Jastrow Ansatz is encoded 
within the expansion coefficients of the 
ground-state wave function with respect to
a complete set of Ising states. In some sense, 
the CCM is found to contain {\it less} correlations 
than the others at `equivalent' levels of 
approximation  (e.g., the CCM LSUB2 
approximation versus Hartree and 
nearest-neighbour Jastrow correlations).
A fuller account of the different 
parametrisations of the ground-state 
wave function within the CCM and CBF 
methods has been given in Ref. \cite{cbf4}.
However, in practice the other methods 
are difficult to extend to approximations 
which contain more than two-body or 
three-body correlations. By contrast, the CCM 
is well-suited to treat such higher-order 
correlations via computational 
techniques, as has been demonstrated here.
Furthermore, the CCM requires no information 
other than the approximation in $S$ and 
$\tilde S$ in order to determine an
approximate ground state of a given system. 
The CBF method, however, may require 
that only a certain subset of all possible
diagrams are summed over (e.g., the HNC/0
approximation). The VQMC approach 
also often requires an intimate knowledge 
of the manner in which the two-body correlations
behave with increasing lattice separation
if all two-body correlations are to be
included. This information may be approximated, 
for example, by use of the results of spin-wave theory.
In any case, it is often necessary to reduce
the minimisation of the variational ground-state 
energy with respect to $N$ parameters to much
fewer parameters. Another potential application 
of all of the methods presented here is 
the use of their ground-state wave functions
as trial or guiding wave functions in (Green 
function or similar) quantum Monte Carlo calculations. 

Finally, encouraged by these results for the
transverse Ising model, we intend to extend
them to other models of interest, such as systems
with higher quantum spin number or those 
with complex crystallographic lattices. 
A further goal is to extend the treatment 
of this and other spin models, via these methods, 
to non-zero temperatures.

\section*{Acknowledgements}

We thank Dr. N.E. Ligterink for his useful and enlightening
discussions. One of us (RFB) gratefully acknowledges a 
research grant from the Engineering and Physical Sciences 
Research Council (EPSRC) of Great Britain. This work has 
also been supported in part by the Deutsche 
Forschungsgemeinschaft (GRK 14, Graduiertenkolleg on 
`Classification of Phase Transitions in Crystalline 
Materials'). One of us (RFB) also acknowledges the support 
of the Isaac Newton Institute for Mathematical Sciences, 
University of Cambridge, during a stay at which the
final version of this paper was written.

\pagebreak

\appendix
\section{Extrapolation of CCM Results}

In this Appendix, we explain how we extrapolate 
a set of LSUB$m$ data points, $\{x_i,y_i\}$, in the limit
$i \rightarrow \infty$ at each value of some 
parameter ($\lambda$) within the Hamiltonian separately. 
Note that the number of data elements to be extrapolated 
is given by the index $p$. The value of $x_i$ is 
now set to be $1/m$ and $y_i$ is set to be the 
corresponding value of an expectation value (for example,
the lattice magnetisation) determined using the CCM 
at this level of approximation at a given value of 
$\lambda$. Note that the value of $m$ must 
increase with increasing index $i$, although $m$ and $i$
do not have to be equal. 

Before the extrapolation procedures are given 
in detail, we define some useful quantities. Firstly, 
the mean value of a set $\{c_i\}$ is denoted by $\overline c$ 
and of a set $\{d_i\}$ is denoted by $\overline d$. Secondly, 
the linear correlation, $R$, of a set of two-dimensional
points, $\{c_i,d_i\}$, is defined by
\begin{equation}
  R \equiv \frac {\sum_{i=1}^p (c_i -\overline c) (d_i -\overline d)}
  {\sqrt{\sum_{i=1}^p (c_i -\overline c)^2} 
  \sqrt{ \sum_{i=1}^p (d_i -\overline d)^2}}
~~ .
\label{app1}
\end{equation}
We are now in a position to outline the the first extrapolation 
procedure. This procedure firstly assumes that the data scales 
with a leading-order ``power-law'' dependence, given by
\begin{equation}
y_i = a + b x_i^{\nu} ~~ .
\label{app2}
\end{equation}
We set $c_i={\rm log}(x_i)$ and $d_i={\rm log}(y_i-a)$, where $\{x_i,y_i\}$
is the LSUB$m$ data set at some fixed value of a parameter within
the Hamiltonian. Hence the best fit of the data set, $\{x_i,y_i\}$, 
to the power-law dependence of Eq. (\ref{app2}) is obtained when 
the absolute value of $R$ is maximised with respect to the variable
$a$. Indeed, we make the assumption that this value of $a$ is 
then taken to be the extrapolated value of the $y_i$ in 
the limit $i\rightarrow \infty$ (in which case, $x_i 
\rightarrow 0$). 

The second extrapolation procedure of the LSUB$m$ data uses 
Pad\'e approximants. This is achieved by firstly assuming that 
the set of data can be modeled by the ratio of two 
polynomials, given by
\begin{equation}
y_i = \frac {\sum_{j=0}^k a_j x_i^j} {1 + \sum_{j=1}^l b_j x_i^j} ~~ .
\label{app3}
\end{equation} 
Note that when $l=0$, this is a simple integral power series. 
This furthermore implies that,
\begin{equation}
a_0 + a_1x_i + a_2 x_i^2 + \cdots + a_k x_i^k
= y_i +(b_1x_i + b_2x_i^2 + \cdots + b_l x_i^l) y_i ~~.
\label{app4}
\end{equation}
We now wish to determine the coefficients $a_j$ and $b_j$ 
in order to find the polynomials in Eq. (\ref{app3}), 
and Eq. (\ref{app4}) is rewritten in terms of a matrix 
given by,
\begin{equation}
\left (
\mbox{
\begin{tabular}{ccccc|cccc}
1, &$x_1$, &$x_1^2$, &$\cdots$, &$x_1^k$  &$x_1y_1$, &$x_1^2y_1$, &$\cdots$, 
&$x_1^ly_1$ \\
1, &$x_2$, &$x_2^2$, &$\cdots$, &$x_2^k$  &$x_2y_2$, &$x_2^2y_2$, &$\cdots$, 
&$x_2^ly_2$ \\
   &$\cdot$&         &$\cdot$   &         &          &$\cdot$     &$\cdot$
&         \\ 
   &$\cdot$&         &$\cdot$   &         &          &$\cdot$     &$\cdot$
&         \\ 
   &$\cdot$&         &$\cdot$   &         &          &$\cdot$     &$\cdot$
&         \\ 
   &$\cdot$&         &$\cdot$   &         &          &$\cdot$     &$\cdot$
&         \\ 
1, &$x_p$, &$x_p^2$, &$\cdots$, &$x_p^k$  &$x_p y_p$, &$x_p^2y_p$, &$\cdots$, 
&$x_p^ly_p$ \\
\end{tabular}
}
\right ) 
\left (
\mbox{
\begin{tabular}{c}
$a_0$   \\
$a_1$   \\
$\cdot$ \\
$a_k$   \\
$-b_1$   \\
$\cdot$ \\
$-b_l$   \\
\end{tabular}
}
\right ) =
\left (
\mbox{
\begin{tabular}{c}
$y_1$ \\
$y_2$ \\
$\cdot$ \\
$\cdot$ \\
$\cdot$ \\
$\cdot$ \\
$y_p$ \\
\end{tabular}
}
\right )
\label{app5}
\end{equation}
The inverse of the matrix in Eq. (\ref{app5}) is now obtained and 
the coefficients $a_j$ and $b_j$ are determined. (Note that
$k+l+1=p$.) However, we also note that because $x_i 
\rightarrow 0$ as $i \rightarrow \infty$, 
$a_0$ gives us the extrapolated value of $\{y_i\}$ in the 
limit $m \rightarrow \infty$. Furthermore, using this method 
with $l=0$ we found that a previous extrapolated result\cite{ccm19} 
of CCM LSUB$m$ data for the {\it sublattice magnetisation} of the 
Heisenberg antiferromagnet on the square lattice was reproduced. 
In this previous calculation\cite{ccm19}, the sublattice magnetisation
was extrapolated in the limit $m \rightarrow \infty$ by fitting 
the LSUB$m$ points (with $m=4,6,8$) to a 
quadratic function in $1/m$ thus giving an extrapolated value of about 
$0.6$.

\begin{figure}
\epsfxsize=10cm
\centerline{\epsffile{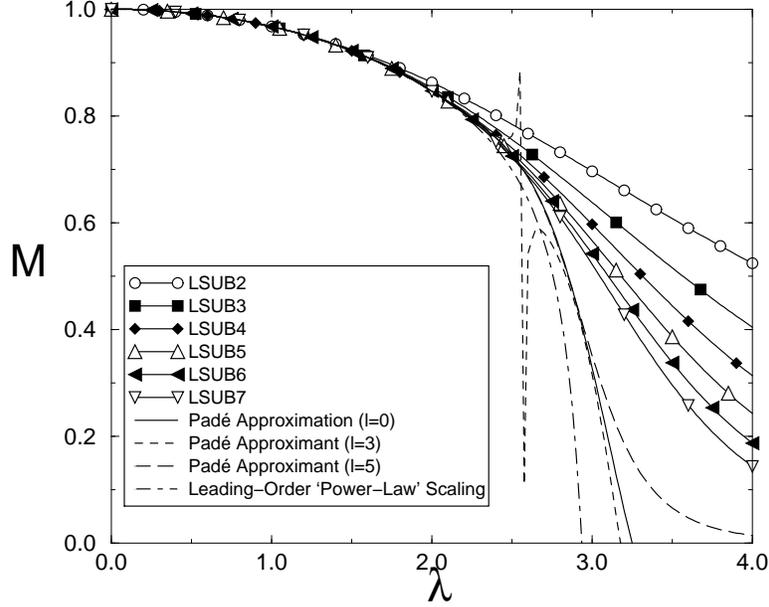}}
\caption{Results for the lattice magnetisation, $M$, of the spin-half
transverse Ising model on the square lattice in the ferromagnetic 
regime for the CCM LSUB$m$ approximation with $m=\{2,3,4,5,6,7\}$.
The extrapolation of the LSUB$m$ results at each separate 
value of $\lambda$ is performed in two ways. The first uses
Pad\'e approximants in order to perform the extrapolation,
and the second assumes a leading-order power-law scaling of the lattice 
magnetisation with $m^{-1}$.}
\label{fig4}
\end{figure}

In this article, we have already plotted extrapolated 
CCM LSUB$m$ results for the lattice magnetisation of 
the square lattice spin-half transverse Ising model
in Fig. \ref{fig2}, and these results were seen to be 
in excellent agreement with those results of CBF 
and VQMC calculations.
However, a further discussion of the extrapolated 
CCM results presented here is also useful in order to illustrate 
the strengths and weaknesses of the extrapolation procedures
outlined in this appendix. We can see from Fig. \ref{fig4} 
below that the results for the Pad\'e approximant 
extrapolation with $l=3$ contains a zero in the 
denominator of Eq. (\ref{app3}) at about $\lambda 
\approx 2.6$ such that the results show a divergence 
for $M$ in Fig. \ref{fig4} which is simply  
an {\it artifact} of the extrapolation procedure. This 
is because an assumption is made as to the scaling of 
the LSUB$m$ data with $1/m$ to some functional form. 
The validity of this assumption is unknown as
no exact scaling laws are known, as yet, for the
behaviour of CCM LSUB$m$ results as functions of 
$m$. However, the empirical 
evidence in Fig. \ref{fig4} suggests that this is a
reasonable assumption over much of the ferromagnetic 
phase, except, of course, for those points at which the 
Pad\'e approximant results demonstrate this `artificial' 
divergence.

\end{document}